\documentclass[prl,twocolumn,showpacs,preprintnumbers,amsmath,amssymb]{revtex4}

\usepackage{amsmath}

\usepackage{graphics}

\usepackage{epsf}

\usepackage{graphicx}
\usepackage{dcolumn}
\usepackage{bm}

\begin{document}

\title{Band structure model of magnetic coupling in semiconductors}

\author{Gustavo M. Dalpian, and Su-Huai Wei}
\affiliation{National Renewable Energy Laboratory, Golden, Colorado 80401, U.S.A}
\author{X. G. Gong}
\affiliation{Department of Physics, Fudan University, Shanghai 200433, China} 
\affiliation{Institute of Solid State Physics, Chinese Academy of Sciences, Hefei 230031, China}
\author{Ant\^onio J. R. da Silva, and A. Fazzio}
\affiliation{Instituto de F\'{\i}sica, Universidade de S\~ao Paulo, CP 66318, 05315-970 S\~ao Paulo,
SP, Brazil}

\date{\today}

\begin{abstract}
We present a unified band structure model to explain magnetic ordering
in Mn-doped semiconductors. This model is based on the $p$-$d$ and $d$-$d$ 
level repulsions between the Mn ions and host elements and can successfully
explain magnetic ordering observed in all Mn doped II-VI and III-V 
semiconductors such as CdTe, GaAs, ZnO, and GaN. This model, therefore, 
provides a simple guideline for future band structure engineering of 
magnetic semiconductors.
\end{abstract}


\maketitle

Mn-doped II-VI and III-V diluted magnetic semiconductors (DMS) have
many unique magneto-optical, magneto-electrical, and magneto-transport properties
that are essential for future-generation spintronic device 
applications \cite{furd88,ohno98,diet00,awsc00,wolf01}.  
These materials also present many interesting behaviors. For example, Mn-doped II-VI
semiconductors generally have antiferromagnetic (AFM) ground state,
whereas Mn doped III-V semiconductors mostly have ferromagnetic
(FM) ground state. Of great interest, some of the systems, such as 
Mn-doped GaN, can be either FM or
AFM, depending on the Mn concentration, carrier density, or
pressure \cite{dalpian04}. It will be, therefore, rewarding to understand the
mechanism of magnetic coupling in these systems in order to design
functional spintronic materials.

\begin{figure}
\epsfxsize 6cm
\centerline{\epsffile{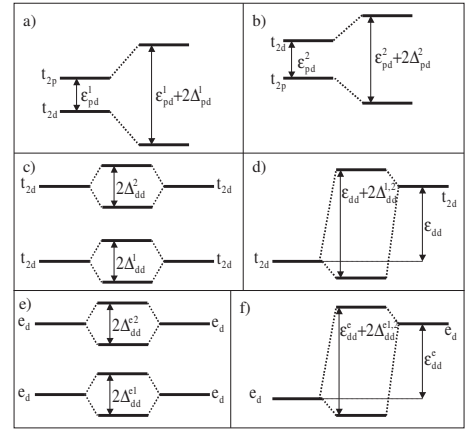}}
\caption{Schematic model for $p$-$d$ and $d$-$d$ couplings. (a)
$p$-$d$ coupling with $t_{2p}$ above $t_{2d}$.  (b) $p$-$d$ coupling
with $t_{2p}$ below $t_{2d}$.  (c) $d$-$d$ coupling between the $t_{2d}$
states with majority spin level coupled to majority spin level and
minority spin level coupled to minority spin level, and (d) $d$-$d$
coupling between the $t_{2d}$ states with majority spin level coupled to
minority spin level. (e) Same as (c), but between the $e_d$ state. (f)
Same as (d), but between the $e_d$ state.
\label{fig1}}
\end{figure}

The mechanism that is responsible for the magnetic coupling in Mn doped III-V
semiconductors has been widely discussed over the last few
years \cite{diet03,litv01,sato03,schilf01}. Several models have been proposed to
explain the phenomena, including the Zener/RKKY and, super-
and double-exchange models. Although these models are quite
successful in explaining some of the systems, they often lack
universality and transparency, and are difficult to compare directly
with {\it ab initio} band structure calculations.  In this paper,
using a band structure approach and level repulsion model, we will
describe a unified picture to explain magnetic ordering in Mn-doped
III-V and II-VI semiconductors and relate them to the previous
models. We show that our model can successfully explain magnetic
ordering observed in all Mn-doped II-VI and III-V semiconductors such
as CdTe, GaAs, ZnO, and GaN. Therefore, it provides simple guidelines
for future band structure engineering of magnetic semiconductors.

Our model is derived from the coupling and level repulsion between the 
magnetic ions $d$ and the host element's $p$ states.  When Mn atoms
are incorporated into a semiconductor, they will introduce $d$ levels
either inside the valence band or above the valence band maximum (VBM)
of the semiconductor, depending on the host material.  For example, for Mn
in GaAs and CdTe, the occupied majority $d$ bands are below the host
VBM, which contains mostly anion $p$ states. The unoccupied minority
$d$ band is above the VBM.  On the other hand, in ZnO and GaN, the
$d$ levels are above the VBM. The introduced $d$ orbitals will
interact with the host $p$
states forming hybrid $pd$ orbitals.
The hybrid orbitals will then also interact with each other.
The hybridization occurs because the Mn $d$ states in a thetrahedral crystal field split into $t_2$ and
$e$ states, whereas the VBM also has the $t_2$ symmetry, with large anion $p$-character.
The two states with the same $t_{2}$ symmetry can couple strongly with each other. 
Furthermore, without spin-orbit (SO) coupling, only the states
with the same spin configuration can couple to each other \cite{so,ze}. Figure 1
shows the possible scenarios for the states that may couple to each other.  
In Fig. 1a, the host $t_{2p}$ state is
above the magnetic ion $t_{2d}$ state, and the coupling leads to a
level repulsion that pushes up the $t_{2p}$ state by $\Delta_{pd}^1$
and pushes down the $t_{2d}$ state by $\Delta_{pd}^1$. In Fig. 1b, the
$t_{2d}$ state is above the $t_{2p}$ state, and they are pushed up and
down respectively by $\Delta_{pd}^2$.  Figure 1c shows the couplings
between the two majority spin $d$ levels
and the two minority spin $d$ levels, which
lead to splittings $2\Delta_{dd}^1$ and $2\Delta_{dd}^2$,
respectively. Figure 1d shows the
coupling between the majority spin $d$ state and the minority spin $d$
state, which pushes up the minority $t_{2d}$ state by
$\Delta_{dd}^{1,2}$ and pushes down the majority spin state by the
same amount. The coupling between the $e_d$ states is similar to that
of between the $t_{2d}$ states, and they are described in Figs. 1e and
1f. In principle, the strenght of the interactions in Fig. 1 depends on the
distance and orientation of the Mn pair. From these figures, we can see that there will be no energy gain
if the two coupled states are fully occupied or fully empty. A
magnetic state can be stabilized only if there are both holes and
electrons in the coupled states.

In general, the $p$-$d$ coupling is larger than the $d$-$d$ coupling
because the host $p$ orbital is more delocalized and Mn $d$ and anion
$p$ are nearest neighbors.  Within the $d$-$d$ coupling, the coupling
between the $e_d$ state is much weaker than the coupling between the
$t_{2d}$ state, because the $e_d$ state is very localized, whereas the $t_{2d}$ state
is less localized due to the coupling with the host $p$ states. 
Because the $d$-$d$ coupling between the $e_d$ state is qualitatively
similar to that of between the $t_{2d}$ state, they are not discussed
explicitly in our model. Furthermore, $\Delta_{dd}^{1,2}$ is smaller
than $\Delta_{dd}^{1}$ when the exchange splitting $\epsilon_{dd}$ is
not zero because the level repulsion is larger if the energy separation
of the two coupling states is small. 

\begin{figure}
\epsfxsize 8.6cm
\centerline{\epsffile{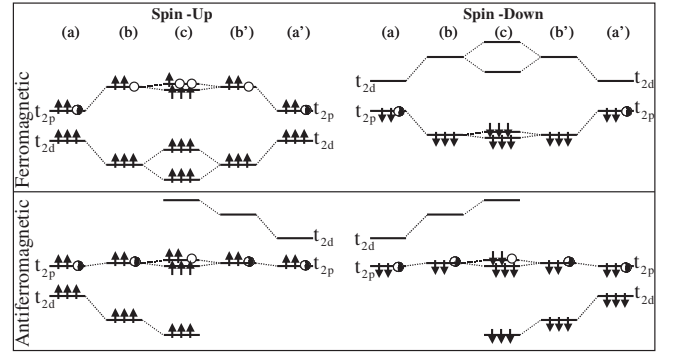}}
\caption{Schematic model showing the position of the $p$ and $d$
levels and level repulsion between them in FM and AFM
configurations. In this case, the VBM is in between the majority and
minority $d$ levels. Note that in (b), (b'), and (c), the states have mixed $pd$ characters.
\label{fig2}}
\end{figure}

With the understanding above, in the following, we will discuss the
mechanism of magnetic ordering observed in Mn doped II-VI and III-V
semiconductors. In the discussion, we will use the two-band coupling model
described in Fig. 1.  First we will discuss the interaction
between the $p$ and $d$ levels, and then the interaction between the
$d$ levels. We will assume the coupling is additive. The small effect of 
coupled $p$-$d$ and $d$-$d$ interactions will be discussed briefly for simplicity, 
but in general, the treatment would not change the results qualitatively.

(i) We discuss first the scenario when the VBM is {\it between} the Mn
majority and minority spin $d$ states.  This is the case for most
Mn-substituted II-VI and III-V systems such as CdMnTe and GaMnAs.  A
schematic plot is shown in Fig. 2, where the level repulsions are
arranged in spin-up and spin-down channels in FM and AFM
configurations.  From (a) to (b) or (a') to (b'), only the $p$-$d$
coupling is turned on.  From (b) and (b') to (c), the $d$-$d$ coupling
is turned on. We first look at the FM configuration.  In the spin-up
channel, the $p$-$d$ coupling pushes the $t_{2p}$ state upward by
$2\Delta_{pd}^1$, whereas it pushes down the $t_{2d}$ state by the
same amount \cite{far}. The net energy gain in this process is
$-2m_h\Delta_{pd}^1$, where $m_h$ is the number of holes.
The $d$-$d$ coupling further splits the $d$ levels by $2\Delta_{dd}^1$.
Because all the $t_{2d}$ majority spin levels are fully occupied,
there is no energy gain in this process. (The $p$-$p$ coupling is
already included in the band structure calculation for the host, and its
effect is the same for FM and AFM cases; therefore, it is not
discussed here.)  In the spin-down channel, the $p$-$d$ coupling lowers
the energy of the occupied $t_{2p}$ spin-down state by
$2\Delta_{pd}^2$. Because there are six electrons in the two $t_{2p}$
states, the net energy gain in this case is
$-12\Delta_{pd}^2$. Therefore, the net energy gain for the FM
configuration is $-2m_h\Delta_{pd}^1 - 12\Delta_{pd}^2$.  In the AFM
configuration, the situation is the same in the spin-up and spin-down
channel. When the $p$-$d$ coupling is turned on, the $t_{2p}$ state is
pushed up by the occupied majority $t_{2d}$ state by $\Delta_{pd}^1$,
but pushed down by the unoccupied minority $t_{2d}$ state by
$\Delta_{pd}^2$. The occupied majority $t_{2d}$ state is pushed down
by $\Delta_{pd}^1$, and it is further pushed down by the unoccupied
minority $t_{2d}$ state by $\Delta_{dd}^{1,2}$. Therefore, the net
energy gain in the AFM configuration is $-m_h\Delta_{pd}^1
-(12-m_h)\Delta_{pd}^2 - 6\Delta_{dd}^{1,2}$. Hence, taking into account
the localization of the host $p$ orbitals around the Mn atoms \cite{far}, the energy
difference between the FM and AFM phase is 
\begin{equation}
\Delta E_{FM-AFM} =
-\alpha m_h(\Delta_{pd}^1+\Delta_{pd}^2)+6\Delta_{dd}^{1,2},
\end{equation}
where $\alpha < 1$ decreases when the hole states become more localized,
and when Mn-Mn distance increases. This result
suggests that (a) the AFM phase is stabilized by the energy
$6\Delta_{dd}^{1,2}$ from coupling between the majority and minority spin $d$ states
(often denoted as
superexchange) \cite{dd}. (b) The FM phase is stabilized with energy
$-m_h(\Delta_{pd}^1+\Delta_{pd}^2)$ which is proportional to the
number of holes and the $p$-$d$ exchange splitting
$(\Delta_{pd}^1+\Delta_{pd}^2)$, as described in the Zener model.
Therefore, to enhance FM coupling, one should increase the hole
carrier density and increase the $p$-$d$ exchange splitting.
However, large $p$-$d$ coupling also leads to large localization of
the hole stae, thus, a balance between $\alpha$ and $\Delta_{pd}$ is needed; (c) For
systems where Mn substitution on the cation site does not introduce
holes (e.g., CdMnTe), the system is always more stable in the AFM
phase. (d) For systems where Mn substitution on the cation site
introduces holes (e.g., GaMnAs, where each Mn on Ga site introduces
one hole), because the $p$-$d$ coupling is larger than the $d$-$d$
coupling, the system in general will have a FM ground state if enough
holes are present in the system. However, when holes are compensated by
donor defects, the system can revert to the AFM ground state.

\begin{figure}
\epsfxsize 8.6cm
\centerline{\epsffile{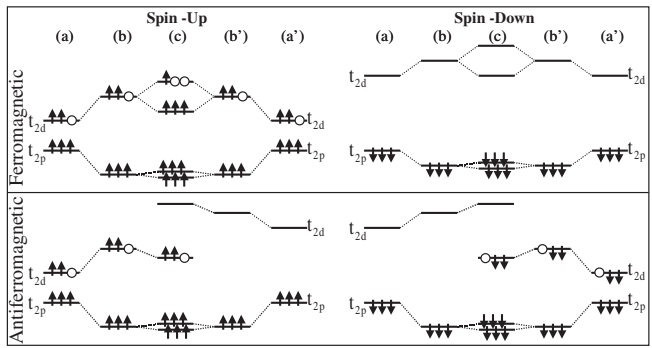}}
\caption{Schematic model showing the position of the $p$ and $d$
levels and level repulsion between them in FM and AFM
configurations. In this case, the Mn $d$ levels are above the VBM.
Note that in (b), (b'), and (c), the states have mixed $pd$ characters.
\label{fig3}}
\end{figure}

\begin{figure}
\epsfxsize 7cm
\centerline{\epsffile{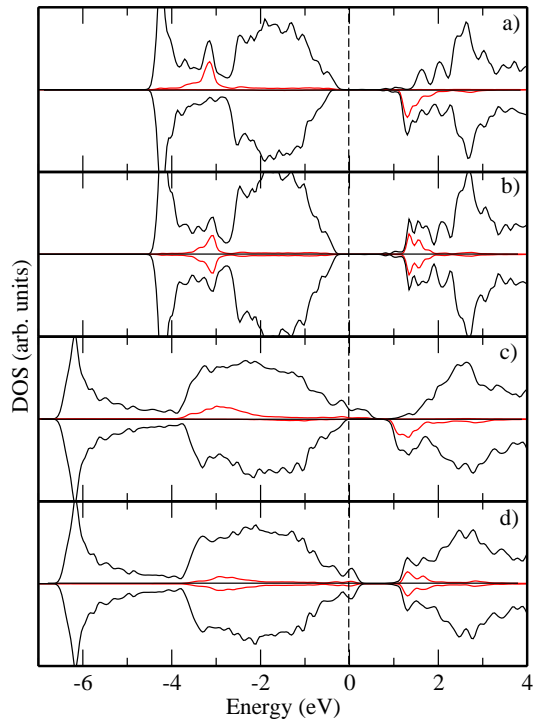}}
\caption{Total (black) and Mn $d$ projected (red) density of states for CdMnTe and GaMnAs:
(a) ferromagnetic CdMnTe; (b) antiferromagnetic CdMnTe; (c) ferromagnetic GaMnAs; and 
(d) antiferromagnetic GaMnAs.
\label{fig4}}
\end{figure}

(ii) In the second scenario we discuss the case when the Mn $d$ state
is {\it above} the VBM of the host.  This is the the case for Mn in
ZnO or in GaN. A schematic plot is shown in Fig. 3. The analysis is
similar to that in the first scenario.  In the FM configuration and
spin-up channel, the system not only gains energy through the $p$-$d$
coupling by $-2m_h\Delta_{pd}^1$, but also through the $d$-$d$
coupling \cite{dd} by $-m_h\Delta_{dd}^1$, which put holes at a high
energy level and electrons at a low energy level.  Here, $m_h \le 3$ is
the number of holes at the $t_{2d}$ level, and for simplicity, we
still use $\Delta_{pd}^1$ to describe the coupling between the VBM and
the majority $t_{2d}$ state. The energy gain in the spin-down channel
is $- 12\Delta_{pd}^2$, thus the total energy gain for the FM
configuration is $-m_h\Delta_{dd}^1 -2m_h\Delta_{pd}^1 -
12\Delta_{pd}^2$. On the other hand, the net energy gain for the AFM
configuration is $-(6-m_h)\Delta_{dd}^{1,2}-2m_h\Delta_{pd}^1 -
12\Delta_{pd}^2$, thus the energy difference between the FM and AFM
phase is
\begin{equation}
\Delta E_{FM-AFM} = -m_h\Delta_{dd}^1+(6-m_h)\Delta_{dd}^{1,2}.
\end{equation}
These results indicate that when the system has holes at the $t_{2d}$-derived
level instead of the $pd$ orbital, (a) the stabilization of the FM or AFM phase
is not directly related to the $p$-$d$ exchange splitting, but is
determined by the $d-d$ coupling \cite{dd} terms $\Delta_{dd}^1$
(often denoted as double exchange) and $\Delta_{dd}^{1,2}$ terms.  (b)
When the number of holes $m_h$ decreases, the AFM phase is stabilized
against the FM phase. When $m_h=0$ (e.g., ZnMnO), the system is always
more stable in the AFM phase.  (c) Whether the system has an FM or AFM
ground state depends not only on hole carrier density, but also on the
relative strength of FM stabilization energy $\Delta_{dd}^1$, and the
AFM stabilization energy $\Delta_{dd}^{1,2}$. $\Delta_{dd}^{1,2}$
increases when the exchange splitting $\epsilon_{dd}$ decreases. This
can be achieved if the charge is transfered from the majority 
to minority spin state (e.g., from majority $t_{2d}$ to
minority $e_d$ state), thus reducing the magnetic moment and exchange
splitting. The majority $t_{2d}$ state can be pushed upward through
increased $p$-$d$ coupling, which can be realized by increasing Mn
concentration or applying pressure. Thus, we predict that for Mn-doped
GaN, when the Mn concentration is high or under pressure, the system
will have an AFM ground state \cite{dalpian04}.

\begin{figure}
\epsfxsize 7cm
\centerline{\epsffile{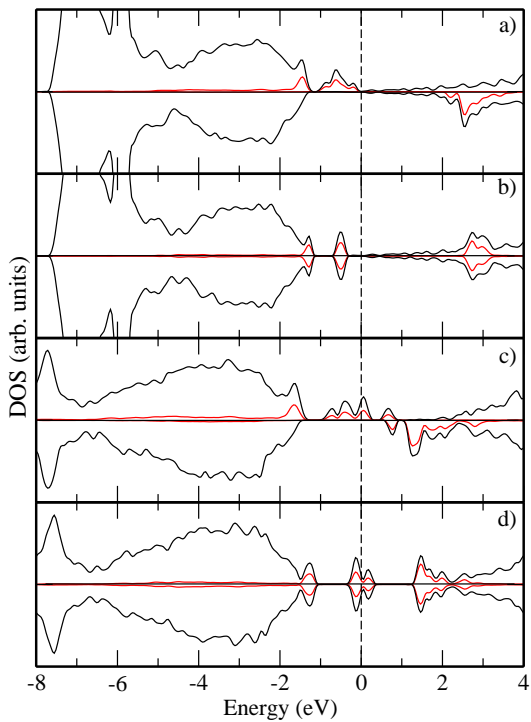}}
\caption{Total (black) and Mn $d$ projected (red) density of states for ZnMnO and GaMnN:
(a) ferromagnetic ZnMnO; (b) antiferromagnetic ZnMnO; (c) ferromagnetic GaMnN; and
(d) antiferromagnetic GaMnN.
\label{fig5}}
\end{figure}

To test our models, we performed first-principles total-energy
calculations for CdMnTe, GaMnAs, ZnMnO, and GaMnN.  The calculations
were performed using an {\it ab initio} plane wave basis code
\cite{vasp}, based on the local spin density functional theory and
using ultrasoft pseudopotentials \cite{uspp}.  For the exchange and
correlation potential, we used the generalized gradient approximation
of Perdew and Wang \cite{pw91}.  The Brillouin zone
integration is performed using the Monkhost-Pack special {\bf k}
points scheme \cite{monk76} of 4x4x4 for the energy differences and
6x6x6 for the DOS, in a 64 atom supercell with two Mn atoms as first fcc neighbors.
Interactions of further neighbors were considered elsewhere \cite{ze}.
We considered the zinc-blende
alloy and assumed that the same results also hold for the alloy in a
wurtzite structure.

Figure 4 shows the total (black) and Mn $d$ projected (red) density of
states for CdMnTe and GaMnAs in the FM and AFM configurations. We see
that these two systems correspond to scenario (i) where the VBM is 
between the majority and minority Mn $d$ states.  For CdMnTe, 
the substitution of Mn for Cd does not introduce holes (Figs.~4a and 4b), therefore, 
according to our model [Eq.~(1)], the system should be AFM. For GaMnAs, holes
are present in the VBM-derived states (Figs.~4c and 4d), therefore, our model predicts that it 
should be FM. Indeed, our directly calculated total energy differences between
FM and AFM configurations for these two systems (Table I) 
agree with those expected from our model.
Figure 5 shows the total (black) and Mn $d$ projected (red) density of states for ZnMnO
and GaMnN in the FM and AFM configurations. We see that these two
systems correspond to scenario (ii) where the VBM is below the Mn $d$
states.  For ZnMnO, no holes are present when Zn is replaced by Mn (Figs.~5a and 5b), 
therefore, according to our model [Eq.~(2)], the system should be AFM. For GaMnN with small
Mn conccentration (6.25\%), holes are created at the Mn $3d$-derived level, and the Mn 
has a high-spin configuration (Figs.~5c and 5d), thus our model predicts that it should have an FM ground state.
Again, our direct calculations shown in Table I agree with what is expected from our model.

\begin{table}
\centering
\caption{Energy difference between FM and AFM configurations for Mn-doped semiconductors ($x=6.25\%$).}
\begin{tabular}{ccc}
\hline\hline
        			&	$\Delta E_{(FM-AFM)}$ (meV)	& Ground state	\\
\hline
Cd$_{1-x}$Mn$_x$Te		&	35			& AF		\\
Ga$_{1-x}$Mn$_x$As		&	-212			& F		\\
Zn$_{1-x}$Mn$_x$O		&	110			& AF		\\
Ga$_{1-x}$Mn$_x$N		&	-225			& F		\\
\hline \hline
\label{table1}
\end{tabular}
\end{table}

In summary, we have proposed a model that can successfully explain the
stabilization of the ferromagnetic or antiferromagnetic phases in 
Mn-doped II-VI and III-V semiconductors.  This simple model is based 
on the $p$-$d$ and $d$-$d$ level repulsions between the Mn ions and 
host states and can be directly related to band parameters. Therefore,
it should be very useful in understanding and engineering diluted 
magnetic semiconductors with desired properties.  A similar
model could also be developed to study other transition metals in
semiconductors, as long as the position of the $d$ levels and the number of
holes induced by it were known. 

The work at NREL is funded by the U.S. Department of Energy, 
Office of Science, Basic Energy Sciences, under Contract No. DE-AC36-99GO10337 to NREL.
XGG is supported by the Nature Science Fundation of China, Chinese
Academy of Science, and National Science Foundation of China.
AF and AJRS thank the support from CNPq and FAPESP.


\begin{thebibliography}{16}
\bibitem[{so()}]{so}
\bibinfo{note}{In a first-order approximation, including spin-orbit coupling
  will not change the magnetic stability qualitatively. See Ref. 12.}
\bibitem[{dd()}]{dd}
\bibinfo{note}{Although the coupling is denoted as $d$-$d$ for simplification,
  due to the strong $p$-$d$ hybridization between the $t_{2d}$ and $t_{2p}$
  states, the states in (b), (b'), and (c) in Figs. 2 and 3 have mixed $pd$
  characters. Because of this hybridization, the so-called $d$-$d$ coupling is
  not exactly a short-range interaction.}
\end{thebibliography}
\end{document}